# Early Results from GLASS-JWST. XXIII. The Transmission of Lyα from UV-faint z ∼ 3–6 Galaxies

Gonzalo Prieto-Lyon[1,2], Charlotte Mason[1,2], Sara Mascia[3,4], Emiliano Merlin[3], Namrata Roy[5], Alaina Henry[5,6],
Guido Roberts-Borsani[7], Takahiro Morishita[8], Xin Wang[9,10,11], Kit Boyett[12,13], Patricia Bolan[14],
Marusa Bradac[14,15], Marco Castellano[3], Amata Mercurio[16,17], Themiya Nanayakkara[18], Diego Paris[3],
Laura Pentericci[3], Claudia Scarlata[19], Michele Trenti[12,13], Tommaso Treu[7], and Eros Vanzella[20]

[1] Cosmic Dawn Center (DAWN), Denmark; gonzalo.prieto@nbi.ku.dk
[2] Niels Bohr Institute, University of Copenhagen, Jagtvej 128, DK-2200 Copenhagen N, Denmark
[3] INAF Osservatorio Astronomico di Roma, Via Frascati 33, I-00078 Monteporzio Catone, Rome, Italy
[4] Dipartimento di Fisica, Università di Roma Tor Vergata, Via della Ricerca Scientifica, 1, I-00133 Roma, Italy
[5] Center for Astrophysical Sciences, Department of Physics and Astronomy, Johns Hopkins University, Baltimore, MD 21218, USA
[6] Space Telescope Science Institute, 3700 San Martin Drive, Baltimore, MD 21218, USA
[7] Department of Physics and Astronomy, University of California, Los Angeles, 430 Portola Plaza, Los Angeles, CA 90095, USA
[8] Infrared Processing and Analysis Center, Caltech, 1200 E. California Blvd., Pasadena, CA 91125, USA
[9] School of Astronomy and Space Science, University of Chinese Academy of Sciences (UCAS), Beijing 100049, People's Republic of China
[10] National Astronomical Observatories, Chinese Academy of Sciences, Beijing 100101, People's Republic of China
[11] Institute for Frontiers in Astronomy and Astrophysics, Beijing Normal University, Beijing 102206, People's Republic of China
[12] School of Physics, University of Melbourne, Parkville 3010, VIC, Australia
[13] ARC Centre of Excellence for All Sky Astrophysics in 3 Dimensions (ASTRO 3D), Australia
[14] Department of Physics and Astronomy, University of California Davis, 1 Shields Avenue, Davis, CA 95616, USA
[15] University of Ljubljana, Department of Mathematics and Physics, Jadranska ulica 19, SI-1000 Ljubljana, Slovenia
[16] Dipartimento di Fisica "E.R. Caianiello," Università Degli Studi di Salerno, Via Giovanni Paolo II, I-84084 Fisciano (SA), Italy
[17] INAF—Osservatorio Astronomico di Capodimonte, Via Moiariello 16, I-80131 Napoli, Italy
[18] Centre for Astrophysics and Supercomputing, Swinburne University of Technology, P.O. Box 218, Hawthorn, VIC 3122, Australia
[19] School of Physics and Astronomy, University of Minnesota, Minneapolis, MN 55455, USA
[20] INAF—OAS, Osservatorio di Astrofisica e Scienza dello Spazio di Bologna, via Gobetti 93/3, I-40129 Bologna, Italy
*Received 2023 April 5; revised 2023 August 30; accepted 2023 September 4; published 2023 October 17*

## Abstract

Lyα emission from galaxies can be used to trace neutral hydrogen in the epoch of reionization, however, there is a degeneracy between the attenuation of Lyα in the intergalactic medium (IGM) and the line profile emitted by the galaxy. Large shifts of Lyα redward of systemic due to scattering in the interstellar medium can boost Lyα transmission in the IGM during reionization. The relationship between the Lyα velocity offset from systemic and other galaxy properties is not well established at high redshift or low luminosities, due to the difficulty of observing emission lines which trace the systemic redshift. Rest-frame optical spectroscopy with JWST/NIRSpec has opened a new window into understanding Lyα at $z > 3$. We present a sample of 12 UV-faint galaxies ($-20 \lesssim M_{\rm UV} \lesssim -16$) at $3 \lesssim z \lesssim 6$, with Lyα velocity offsets, $\Delta v_{\rm Lyα}$, measured from the Very Large Telescope/MUSE and JWST/NIRSpec from the GLASS-JWST Early Release Program. We find a median $\Delta v_{\rm Lyα}$ of 205 km s$^{-1}$ and standard deviation of 75 km s$^{-1}$, compared to 320 and 170 km s$^{-1}$, respectively, for $M_{\rm UV} < -20$ galaxies in the literature. Our new sample demonstrates the previously observed trend of decreasing Lyα velocity offset with decreasing UV luminosity and optical line velocity dispersion, which extends to $M_{\rm UV} \gtrsim -20$, consistent with a picture where the Lyα profile is shaped by gas close to the systemic redshift. Our results imply that during reionization Lyα from UV-faint galaxies will be preferentially attenuated, but that detecting Lyα with low $\Delta v_{\rm Lyα}$ can be an indicator of large ionized bubbles.

*Unified Astronomy Thesaurus concepts:* Emission line galaxies (459); Galaxies (573); Lyman-alpha galaxies (978); Galaxy evolution (594); High-redshift galaxies (734)

## 1. Introduction

Lyα (1216 Å) emission from astrophysical sources has long been used as a tracer of neutral gas in the intergalactic medium (IGM; e.g., Gunn & Peterson 1965; Miralda-Escude 1998). At $z \gtrsim 6$, the declining strength of Lyα emission from galaxies has been used to infer the timing of reionization (e.g., Ouchi et al. 2010; Stark et al. 2010; Treu et al. 2012; Mesinger et al. 2015; Mason et al. 2018a; Morales et al. 2021; Bolan et al. 2022), complementary to measurements from the cosmic microwave background (CMB) electron scattering optical depth (Planck Collaboration et al. 2020) and the optical depth in the Lyα forest of quasars (e.g., Fan et al. 2006; McGreer et al. 2015; Lu et al. 2020; Qin et al. 2021).

However, there is a degeneracy between the strength of the damping wing absorption—due to the abundance of neutral hydrogen in the IGM—and the Lyα emission line profile emerging from the galaxy—set by scattering by neutral hydrogen within the interstellar and circumgalactic media (ISM and CGM, respectively; e.g., Neufeld 1991). If a Lyα emission line emerges from a galaxy with its flux profile redshifted to $\gtrsim$300 km s$^{-1}$ from the systemic velocity of the galaxy, these photons will experience a greatly reduced optical depth in the IGM and thus have high Lyα transmission, even in a highly neutral IGM (e.g., Dijkstra et al. 2011; Mason et al. 2018a; Mason & Gronke 2020; Endsley et al. 2022). Indeed, high rates of Lyα detection that have been reported in $z > 7.5$ luminous galaxies (Oesch et al. 2015; Zitrin et al. 2015; Roberts-Borsani et al. 2016; Stark et al. 2017) could be partially explained if substantial H I ISM reservoirs in these systems strongly scatter Lyα, and/or if strong outflows are more common in UV-bright galaxies, such that Lyα emerges from the galaxy at highly redshifted velocities (Mason et al. 2018b; Endsley et al. 2022; Tang et al. 2023). Nonetheless, even if Lyα transmission is boosted by high-velocity offsets, the resulting broadening of the profile is likely accompanied by a weakening of the line, reducing its detectability (Verhamme et al. 2018).

Until the advent of JWST, measurements of Lyα velocity offsets have only been possible for statistical samples at $z \lesssim 2$, where rest-frame optical emission is visible from the ground for Lyman-break galaxies (LBGs; e.g., Erb et al. 2014; Steidel et al. 2014); for local analogs, strong line emitters at $z \sim$ 0.3–0.4, with detailed Hubble Space Telescope (HST)/Cosmic Origins Spectrograph (COS) spectroscopy (e.g., Henry et al. 2015; Yang et al. 2017b; Hayes et al. 2023); and for smaller samples of UV-luminous sources with bright [C II] emission lines visible with Atacama Large Millimeter/submillimeter Array (ALMA; Willott et al. 2015; Inoue et al. 2016; Pentericci et al. 2016; Bradač et al. 2017; Cassata et al. 2020; Endsley et al. 2022), or the highly ionized UV line C III] (Stark et al. 2015, 2017), or in stacks of absorption line spectra of Lyα emitters (LAEs; Muzahid et al. 2020). Thus, reionization inferences have relied on extrapolating $z < 2$ samples to higher redshifts and direct measurements for only a handful of galaxies in the epoch of reionization (e.g., Mason et al. 2018a). JWST's near-infrared (NIR) spectroscopy opens the gates to accurate measurements of the Lyα velocity offset and its evolution with galaxy properties, which will allow us to break the degeneracy between ISM scattering and IGM opacity.

In this paper we present a first look at the Lyα velocity offsets, $\Delta v_{Ly\alpha}$, of $z > 3$ UV-faint galaxies, made possible for the first time by JWST. We use JWST/NIRSpec (Jakobsen et al. 2022) spectra of rest-frame optical emission lines, obtained as part of the GLASS-JWST-ERS program (Treu et al. 2022), to measure the Lyα velocity offsets of 12 $M_{UV} \gtrsim -20$ galaxies at $z \sim 3–6$, with lower luminosity and at higher redshifts than previously possible. We explore empirical trends between $\Delta v_{Ly\alpha}$ and other galaxy properties, to explore the physical mechanism behind $\Delta v_{Ly\alpha}$, and we discuss our results in the context of Lyα observations during the epoch of reionization.

The paper is structured as follows. In Section 2 we describe the photometric and spectroscopic data used in our study. In Section 3 we describe how we measure the Lyα velocity offsets for our sample and other relevant properties for the galaxies. We describe our results and comparison to the literature in Section 4. We discuss our results in Section 5 and state our conclusions in Section 6.

We assume a flat Λ cold dark matter (ΛCDM) cosmology with $\Omega_m = 0.3$, $\Omega_\Lambda = 0.7$, and $h = 0.7$; all magnitudes are in the AB system; and all distances are proper unless specified otherwise.

## 2. Data

In this section we describe the data and sample selection used for our study. All observations are made of the lensing cluster A2744.

### 2.1. VLT/MUSE Spectroscopy

LAEs at $2.9 < z < 6.7$ were selected from Very Large Telescope (VLT)/MUSE spectroscopy of A2744 performed by the ESO program 094.A-0115 (Mahler et al. 2018; Richard et al. 2021). These data consist of four 1 arcminute$^2$ regions centered on the A2744 cluster core. These quadrants have observing times of 3.5, 4, 4, and 5 hr respectively, with two extra hours of observation overlapping at the center of the cluster. The MUSE data have a median emission line flux 1σ uncertainty of $3.6 \times 10^{-19}$ erg s$^{-1}$ cm$^{-2}$, translating to an equivalent width (EW) 5σ limit of ∼4–30 Å for an $M_{UV} = -19$ galaxy at $z \sim 3-7$.

The MUSE program applied three complementary detection methods: (i) forced extraction in the location of $m_{AB} \lesssim 30$ sources detected in Hubble Frontier Fields imaging (HST-GO/DD-13495; Lotz et al. 2017), using an extraction aperture corresponding to the SExtractor segmentation map convolved with the MUSE point-spread function; (ii) detection of emission lines via narrow-band filtering of the MUSE cube; and (iii) manual extractions of a field sources found through visual inspection, e.g., multiply imaged systems. In the following, we will compare MUSE-selected LAEs to, mostly, Lyman-break-selected samples. It is known that LAEs may be the extreme end of the LBG population (e.g., Dijkstra & Wyithe 2012; Morales et al. 2021), undergoing very recent star formation, and are more likely to be low mass, less dusty, and show lower Lyα velocity offsets than Lyman-break-selected samples (e.g., Hashimoto et al. 2013; Shibuya et al. 2014).

In this study, we are most interested in understanding the expected properties of Lyα emission from Lyman-break-selected galaxies at $z \gtrsim 6$, during the epoch of reionization. At $z \sim 6$ a much larger fraction (∼30%−60%) of UV-faint galaxies are detected with high Lyα EW (>25 Å; e.g., Stark et al. 2011; Cassata et al. 2015; De Barros et al. 2017; Fuller et al. 2020) than at $z \lesssim 3$ (∼10%; e.g., Cassata et al. 2015). Furthermore, due to both an expected reduction in dust attenuation (as evidenced by steepening UV slopes, β; e.g., Bouwens et al. 2014) and an increase in specific star formation rates (e.g., Stark et al. 2013; Endsley et al. 2023) we should expect strong Lyα emission to be even more prevalent at $z > 6$. Therefore, we assume that the MUSE LAE sample is a suitable analog of UV-faint $z \gtrsim 6$ LBGs.

To test this assumption we compare with the sample of Fuller et al. (2020) who obtained deep Lyα spectroscopy of lensed UV-faint LBGs at $z \sim 6$, with that of Richard et al. (2021; the origin catalog for our Lyα-selected galaxies). We select sources in the Fuller et al. (2020) catalog with $5.5 < z < 6.5$ (using photometric redshift for nondetections, where we only include sources with $P(5.5 < z_{phot} < 6.5) > 0.6$ to minimize contamination by low-redshift interlopers) and $M_{UV}$ in the same range as our sample. Nondetections for this subset of the Fuller et al. (2020) catalog have a median upper limit EW $\lesssim$ 42 Å. The Richard et al. (2021) sample has a slightly bluer UV slope by $\Delta\beta = 0.4$. This means that we are possibly studying a population slightly younger and/or less dust attenuated than galaxies selected by their Lyman break. However, we find that the Lyα EW distributions have similar medians, Lyα EW = 36 Å for our parent MUSE sample, and the Fuller et al. (2020) sample has a median Lyα EW = 42 Å if all nondetections are treated as having a Lyα EW uniformly distributed between 0 and their EW$_{upperlim}$. Performing 10,000 iterations, and doing a Kolmogorov–Smirnov (K-S) test of each, we find a median 80% chance that the underlying distributions are statistically the same. We conclude it is highly likely that the underlying EW distributions of our parent sample are representative of $5.5 < z < 6.5$ UV-faint LBG-selected galaxies.

We further compare the Lyα EW of Richard et al. (2021) with all LBGs targeted by the GLASS-NIRSpec program between $3 < z_{phot} < 7$. To form the subsample of LBGs, we apply a color criterion to their HST photometry (Bouwens & Illingworth 2015; Bouwens et al. 2016), which has a median $z \sim 4$. We match the resulting 48 NIRSpec LBGs to Richard et al. (2021), obtain Lyα fluxes, and compute their EWs. For 38 galaxies we find no Lyα counterpart, and use the 5σ flux limit of the MUSE observations. As in the previous paragraph, we treat upper limits as a uniform distribution and iterate 10,000 times performing a K-S test between each distribution. Including the nondetections, we find it unlikely ($p < 5\%$) that the LBG-selected sample has the same Lyα EW distribution as our MUSE LAE parent sample. In summary, we conclude that our parent sample (Richard et al. 2021) is likely biased toward high EWs compared to $z \sim 4$ LBG-selected galaxies, but that it is comparable to UV-faint LBG-selected sources at the end of reionization, $z \sim 6$.

We note that it is possible that the Lyα fluxes reported by Richard et al. (2021) are underestimated, since they do not consider the spatially extended Lyα emission of the galaxies for their Lyα extractions. Measurements including the spatially extended Lyα flux for MUSE LAEs with NIRSpec observations in the GLASS-ERS survey are presented by Roy et al. (2023). In this we use the fluxes as reported by Richard et al. (2021), which we consider more similar measurements to those obtained by slit spectrographs, which is how the majority of our comparison samples have been observed.[21]

### 2.2. JWST NIRSpec Spectroscopy

Near-infrared spectroscopic data in A2744 were obtained from the GLASS-JWST Early Release Science Program (PID 1324; Treu et al. 2022), observed on 2022 November 10. Three grating configurations were used, G140H/F100LP, G235H/F170LP, and G395H/F290LP, each with a 4.9 hr exposure time. This allows observations spanning the wavelength range 0.81–5.14 μm. It is possible to observe Hα up to $z \sim 6.8$ and [O III] λ5007 up to $z \sim 9.2$. The high-resolution, $R \sim 2700$ (corresponding to $\sigma \sim 50$ km s$^{-1}$), allows for precise measurements of the spectral properties of the emission lines: most important for us are the line centroid and dispersion.

Details of the reduction process are given by Roberts-Borsani et al. (2022) and Morishita et al. (2023). Briefly, we use the STScI JWST pipeline (ver.1.8.2)[22] for Level 1 data products, and the `msaexp`[23] Python package for Level 2 and 3 data products. We extract the 1D spectrum following Morishita et al. (2023) and use the `msaexp` package to optimize our extraction using an inverse-variance-weighted kernel following Mascia et al. (2023). The kernel is derived by summing the 2D spectrum along the dispersion axis and fitting the resulting signal along the spatial axis with a Gaussian profile ($\sigma \sim 0.''4$). This 2D kernel is then used to extract the 1D spectrum along the dispersion axis.

MUSE LAEs from the Mahler et al. (2018) and Richard et al. (2021) catalog were included in the GLASS-NIRSpec MSA. The MSA target selection is described by Treu et al. (2022); briefly, $z > 5$ and spectroscopically confirmed sources were prioritized. Therefore, the sample of MUSE LAEs observed is not guaranteed to be magnitude complete, but provides a qualitative comparison sample to UV-bright literature samples galaxies with measured velocity offsets.

A total of 17 MUSE LAEs were observed with NIRSpec in the GLASS-JWST-ERS program. We detect rest-frame optical emission lines for 12 out of the 17 sources in the NIRSpec spectra, which we use for our work. From the remaining five sources we did not detect any optical emission lines (see also Mascia et al. 2023). The lack of detection is not surprising: the five sources are faint ($m_{F150W} \sim 27.5$) and were selected from a single faint Lyα line detection. The properties of our sample are presented in Table 1.

We check if our 12 galaxy subsample of Richard et al. (2021) is representative of the complete sample. Following the methods in Section 2.1, we extend our statistical K-S test to our 12 galaxy subsample and compare them to Richard et al. (2021). We find a $\sim70\%$ probability that the EW$_{Ly\alpha}$ distributions of these two are the same. If we compare the 12 galaxy sample with Fuller et al. (2020; $5.5 < z < 6.5$) and the LBGs targeted by GLASS-NIRSpec ($z \sim 4$), we arrive to the same conclusion of Section 2.1: our 12 galaxy sample behaves similar to it is parent sample and is representative of UV-faint LBGs at the end of reionization.

### 2.3. Photometric catalogs

We use the photometric catalogs of A2744 from Paris et al. (2023). The catalogs include HST/Advanced Camera for Surveys (ACS) and WFC3/IR from the Hubble Frontier Fields project (HST-GO/DD-13495; Lotz et al. 2017); photometric bands include F435W, F606W, F814W, F105W, F125W, F140W, and F160W. These catalogs also include JWST/NIRCam data from the UNCOVER collaboration (GO-2561; Bezanson et al. 2022), including photometry from the following seven JWST/NIRCam filters: F115W, F150W, F200W, F277W, F356W, F410M, and F444W. The catalogs of Paris et al. (2023) use F444W as the detection band but as we require the photometry to measure the UV rest-frame continuum of our sources, we only use HST photometry in this work as our sample are all at $z < 6$.

### 2.4. Gravitational Lensing Models

To account for the magnification caused by the strong lensing of the A2744 cluster, we use the lensing maps developed by Bergamini et al. (2023), which can be accessed in their online tool.[24] We do not find any galaxies near the critical curves, with all sources having magnifications between $\mu \sim 2$ and 7, with less than 10% uncertainties.

---

[21] We note that this assumption may not be valid if Lyα is significantly offset from the UV continuum, but recent work shows this is not likely to be the significant at the redshifts and magnitude range we are considering (Hoag et al. 2019b; Lemaux et al. 2021).
[22] https://github.com/spacetelescope/jwst
[23] https://github.com/gbrammer/msaexp
[24] http://bazinga.fe.infn.it:5007/SLOT

Table 1
Lyα and Rest-frame Optical Properties of z ∼ 3–6 Galaxies in GLASS-ERS Used in This Work

| ID | R.A. | Decl. | $z_{\rm sys}$ | $z^*_{\rm Ly\alpha}$ | $\mu$ | $M_{\rm UV}$ * | $\beta$ | Lyα EW (Å) * | $\Delta v_{\rm Ly\alpha}$ (km s$^{-1}$) | $\sigma_{\rm opt}$ (km s$^{-1}$) |
|---|---|---|---|---|---|---|---|---|---|---|
| 90155 | 3.60687 | −30.38557 | 2.9402 | 2.9430 | 1.9 ± 0.1 | −16.6 ± 0.1 | −2.2 ± 0.2 | 59 ± 12 | 207 ± 83 | <44 |
| 80053 | 3.57497 | −30.39678 | 3.1287 | 3.1327 | 3.7 ± 0.1 | −19.2 ± 0.1 | −2.2 ± 0.1 | 11 ± 1 | 287 ± 11 | 51 ± 3 |
| 80113 | 3.60465 | −30.39223 | 3.4724 | 3.4782 | 2.0 ± 0.1 | −17.6 ± 0.1 | −2.0 ± 0.1 | 26 ± 5 | 390 ± 142 | <42 |
| 80027 | 3.56929 | −30.40963 | 3.5796 | 3.5831 | 2.0 ± 0.1 | −19.4 ± 0.1 | −2.4 ± 0.1 | 7 ± 1 | 235 ± 38 | 47 ± 0 |
| 80029 | 3.60318 | −30.41571 | 3.9509 | 3.9540 | 2.8 ± 0.1 | −18.9 ± 0.1 | −2.0 ± 0.1 | 7 ± 1 | 192 ± 78 | 48 ± 1 |
| 80013 | 3.56934 | −30.40873 | 4.0428 | 4.0454 | 2.0 ± 0.1 | −18.3 ± 0.1 | −2.3 ± 0.1 | 47 ± 5 | 162 ± 19 | 50 ± 3 |
| 80085 | 3.57435 | −30.41253 | 4.7246 | 4.7290 | 2.1 ± 0.1 | −15.4 ± 0.4 | −2.5 ± 1.0 | 85 ± 11 | 230 ± 56 | 35 ± 2 |
| 80070 | 3.58232 | −30.38765 | 4.7968 | 4.8016 | 5.3 ± 0.2 | −18.1 ± 0.1 | −1.8 ± 0.1 | 43 ± 3 | 245 ± 9 | 58 ± 1 |
| 70017 | 3.60666 | −30.39328 | 5.1864 | 5.1892 | 2.0 ± 0.1 | <−16.2 | ⋯ | >25 | 132 ± 105 | <30 |
| 70018 | 3.58790 | −30.41159 | 5.2824 | 5.2843 | 6.8 ± 0.2 | −18.2 ± 0.1 | −2.2 ± 0.1 | 8 ± 1 | 83 ± 42 | 23 ± 4 |
| 70022 | 3.57113 | −30.39295 | 5.4292 | 5.4335 | 3.5 ± 0.1 | <−15.7 | ⋯ | >21 | 199 ± 150 | <41 |
| 70003 | 3.57585 | −30.38929 | 5.6180 | 5.6210 | 4.6 ± 0.1 | −16.3 ± 0.2 | −3.7 ± 1.2 | 104 ± 19 | 144 ± 25 | 40 ± 10 |

**Notes.** For nondetections, 5σ upper limits are shown.

## 3. Methods

In the following section we present our methods to retrieve the UV magnitudes of our sample (Section 3.1), Lyα EWs and velocity offsets $\Delta v_{\rm Ly\alpha}$, and the dispersion of optical emission lines (Section 3.2). All resulting values are shown in Table 1.

### 3.1. Measuring $M_{\rm UV}$

To measure $M_{\rm UV}$ (absolute magnitude at 1500 Å) and $\beta$ slope we fit a power-law model $f \propto \lambda^\beta$ to the photometry (e.g., Rogers et al. 2013). We use all the photometric bands that have their response cutoffs in the UV continuum between 1216 and 3000 Å. We fit the model using Markov Chain Monte Carlo sampling with emcee (Foreman-Mackey et al. 2013). Flat priors are used: $-25 < M_{\rm UV} < -12$ and $-4 < \beta < 1$, which cover the expected ranges (e.g., Bouwens et al. 2014). We correct the UV magnitudes for gravitational lensing magnification using the lens model of Bergamini et al. (2023). We take $M_{\rm UV}$, $\beta$, and the corresponding 1σ errors from the resulting posterior distributions of the fitted power laws.

Three sources were not detected in the Paris et al. (2023) catalogs, due to their extremely low continuum flux. In these cases we perform forced photometry at the coordinates expected from the Lyα (MUSE) observations, with a circular aperture of 0″.28 corresponding to the 2 × FWHM aperture of the F444W point-spread function. One of these galaxies was (ID: 80085) was detected in the forced extraction, but not detected in the Paris et al. (2023) catalogs as it is close to a nondeblended source. We report the other two galaxies' (IDs 70017 and 70022) $M_{\rm UV}$ measurements as 5σ upper limits in Table 1.

The median $M_{\rm UV}$ and 1σ scatter of our sample, corrected for magnification, is $-17.8 \pm 1.4$.

### 3.2. Lyα EWs and Velocity Offsets

To measure the Lyα EWs of these 12 galaxies we use the Lyα fluxes reported in the public catalogs of Richard et al. (2021), and the continuum flux density as inferred from the HST photometry described in Section 2.3. We use the posteriors of $M_{\rm UV}$ and $\beta$ obtained in Section 2.3 to infer the posterior distribution of flux densities at 1250 Å. The median EW and 1σ scatter of our sample is $34 \pm 21$ Å.

We measure Lyα velocity offsets as follows:

$$\Delta v_{\rm Ly\alpha} = c \left( \frac{z_{\rm Ly\alpha} - z_{\rm sys}}{1 + z_{\rm sys}} \right) \qquad (1)$$

where $z_{\rm sys}$ is the systemic redshift measured from rest-frame optical emission lines with NIRSpec and $z_{\rm Ly\alpha}$ is the redshift measured from Lyα, corrected to vacuum wavelength, as reported in the MUSE catalog (Richard et al. 2021). For Lyα, the redshift is measured at the peak of the red peak, if the line is double peaked (J. Richard 2023, private communication).

For sources at z < 6.5 we use Hα and [O III] λ5007 emission in NIRSpec to measure velocity offsets. For sources at 6.5 < z < 6.7 only [O III] λ5007 is visible, due to the wavelength coverage. We measure the redshifts of Hα and [O III] λ5007 by performing a Gaussian profile fit using Markov Chain Monte Carlo sampling with emcee (Foreman-Mackey et al. 2013), and recovering the posterior distribution of the central wavelength of the lines and the velocity dispersion ($\sigma_{\rm opt}$) of the line profile. We repeat this process separately for Hα and [O III] λ5007, and take $\sigma_{\rm opt}$ as the average. We include any redshift difference between the lines in the $\Delta v_{\rm Ly\alpha}$ error. We propagate the errors in the central wavelength of the optical lines, and the Lyα central wavelength measured by Richard et al. (2021) for our measurement of $\Delta v_{\rm Ly\alpha}$. Uncertainties in all cases are dominated by the precision of the MUSE/VLT measurements of the Lyα centroid, especially in low signal-to-noise ratio sources. The precision of the systemic redshifts obtained from the optical emission lines (JWST/NIRSpec) is extremely high, with errors ∼1–8 km s$^{-1}$.

Discrepancies between ground-based and JWST/NIRSpec redshifts of the same observed emission lines at >1 μm have been reported (Larson et al. 2023; Tang et al. 2023). This is likely due to the ubiquity of atmospheric skylines in ground-based spectra in the near-infrared, which introduce considerable uncertainty in line centroid measurements, though there may also be uncertainties in the wavelength calibration. To verify the consistency of the MUSE and NIRSpec wavelength calibrations we check three galaxies in the A2744 field where it is possible to compare the redshifts measured from nebular lines which trace systemic redshifts both with MUSE and NIRSpec. For the systemic redshifts we use C III] λ1909 detections for MUSE (Richard et al. 2021), and Hα and/or [O III] for NIRSpec. We find a negligible difference in the redshifts between the two instruments, with a median offset of ∼30 km s$^{-1}$, which is well within the uncertainties of our $\Delta v_{Ly\alpha}$ measurements, and below the instrumental resolution. We also find one galaxy observed by GLASS-ERS which had a previous systemic redshift measurement from VLT/KMOS from the [O II] λλ3726, 3729 doublet (Mason et al. 2017) and find similarly good agreement. Therefore, it is unlikely that any systematic differences between the redshifts measured from space and ground-based telescopes bias our results.

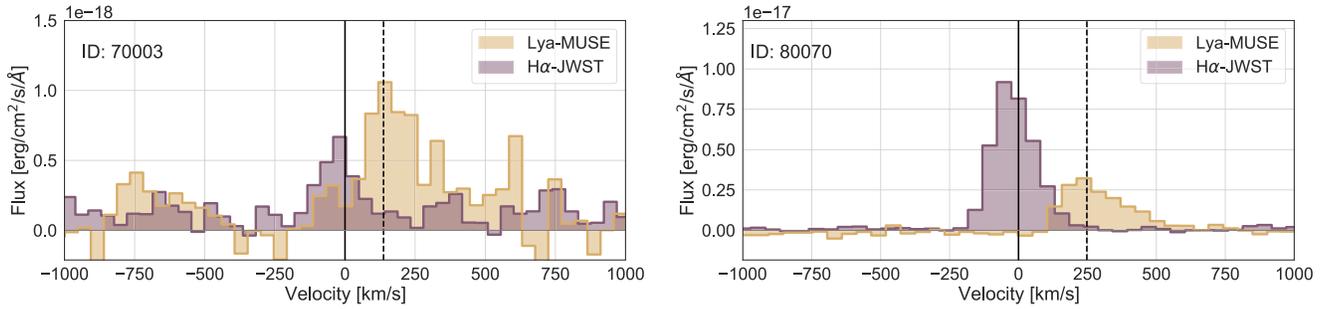

**Figure 1.** Example spectra from this sample, with the velocity axis centered at the systemic redshift given by Hα. We show galaxies 70003 and 80070, which have low and high $\Delta v_{Ly\alpha}$, respectively. Spectra are from JWST/NIRSpec Hα (purple) and VLT/MUSE Lyα (yellow). Vertical lines mark the Hα (solid) and Lyα (dashed) peaks.

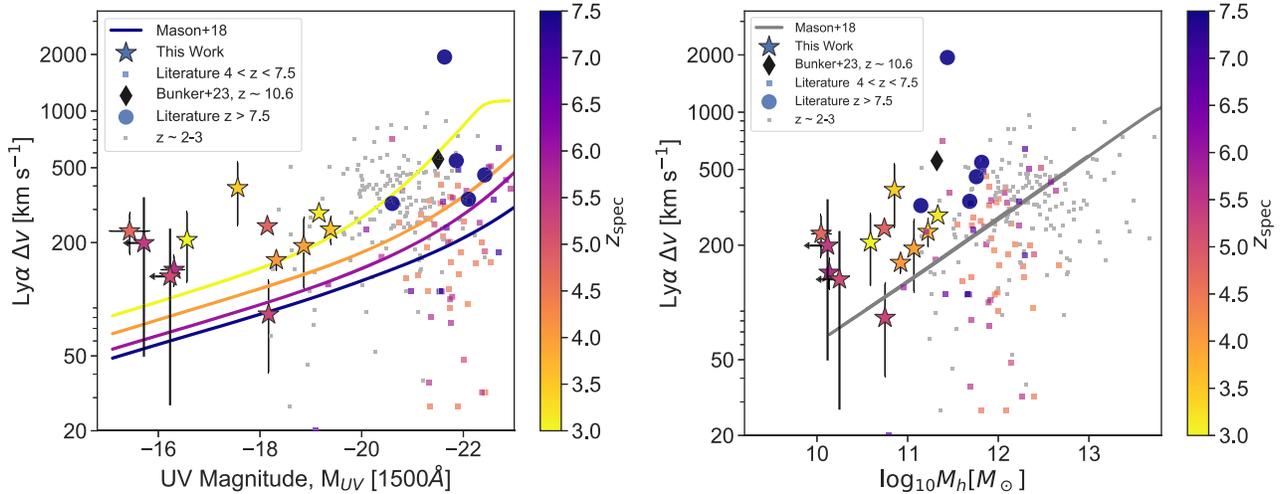

**Figure 2.** Lyα velocity offset vs. UV magnitude and implied halo mass, using the mapping derived in the UV luminosity function model of Mason et al. (2015), color coded by redshift. Our new sample is shown as colored stars. For comparison, we show values from the literature at $z \sim 2$–3 (Erb et al. 2014; Steidel et al. 2014; gray squares), at $4 < z < 7.5$ (Bradač et al. 2017; Stark et al. 2015; Willott et al. 2015; Inoue et al. 2016; Pentericci et al. 2016; Stark et al. 2017; Mainali et al. 2018; Cassata et al. 2020; Endsley et al. 2022; colored squares), at $z > 7.5$ (Tang et al. 2023; blue circles), and GN-z11 (Bunker et al. 2023; black diamond). We also show the model of Mason et al. (2018a) which was derived from $z \sim 2$ literature samples. The new GLASS-ERS sample extends this measurement to galaxies ∼2 mag fainter than previously possible. $\Delta v_{Ly\alpha}$ error bars are dominated by the uncertainty of the centroid in the MUSE Lyα emission observations.

## 4. Results

In the following section we present our results and look for evidence of trends that can explain the mechanisms behind the shift of Lyα emission from the systemic redshift. As described above, our sample is not magnitude complete so our comparison to the literature is qualitative and the sample selection is different from literature samples which are LBG selected (and we expect LAEs to be typically lower mass and less dusty than LBGs; e.g., Kornei et al. 2010; Shibuya et al. 2014). Nevertheless, we find our sample is representative of $z \sim 6$ UV-faint LBGs, of interest during reionization. Our sample allows us to quantify Lyα velocity offsets in UV-faint galaxies for the first time, and investigate the trends by comparing with UV-bright galaxies from the literature.

Our measured Lyα velocity offsets, velocity dispersion, rest-frame EW Lyα, and other observed quantities are presented in Table 1. In Figure 1 we show examples of the spectra observed with JWST/NIRSpec and legacy spectra from VLT/MUSE extracted by Roy et al. (2023).

### 4.1. $\Delta v_{Ly\alpha}$ and UV Magnitude

In Figure 2 we show our measured Lyα velocity offsets versus the UV magnitude of the sample, as well as previous measurements from the literature at $z \sim 2$–3 (Erb et al. 2014; Steidel et al. 2014); $4 < z < 7.5$ (Stark et al. 2015, 2017; Willott et al. 2015; Inoue et al. 2016; Pentericci et al. 2016; Bradač et al. 2017; Mainali et al. 2018; Cassata et al. 2020; Endsley et al. 2022); and $z > 7.5$ (Stark et al. 2017; Tang et al. 2023). Thanks to gravitational lensing and the sensitivity and wavelength coverage of JWST/NIRSpec to detect rest-frame optical lines, we are able to reach much fainter UV magnitudes than in previous works, down to $M_{UV} \sim -16$. The median $M_{UV}$ with $1\sigma$ scatter of our sample is $-17.8 \pm 1.4$ and $\Delta v_{Ly\alpha} = 205 \pm 75$ km s$^{-1}$. This is consistent with the median $\Delta v_{Ly\alpha}$ measured for a stack of $z \approx 3.3$ LAEs from the MUSEQuBES survey (their median $\Delta v_{Ly\alpha} = 171 \pm 8$ km s$^{-1}$; Muzahid et al. 2020).

The left panel of Figure 2 shows a clear trend of increasing $\Delta v_{\mathrm{Ly}\alpha}$ with increasing luminosity. Our new results push these measurements to the UV-faint and low-$\Delta v_{\mathrm{Ly}\alpha}$ end, demonstrating that the trend seen in the literature at brighter luminosities extends to lower luminosities. We compare our observations with the semiempirical model of Mason et al. (2018a), which assumes an underlying relation between Ly$\alpha$ velocity offsets and halo mass, which is independent of redshift. This assumes Ly$\alpha$ scattering is predominantly determined by the total mass of the galaxy, which in turn is assumed to be proportional to the column density of neutral hydrogen, which is predicted to be a key factor in shaping Ly$\alpha$ line profiles (e.g., Neufeld 1990; Verhamme et al. 2006, 2008; Gronke et al. 2016; Kakiichi & Gronke 2021). In this model, the redshift evolution of $\Delta v_{\mathrm{Ly}\alpha}$ arises only from the predicted evolution in the UV luminosity–halo mass relation, whereby galaxies with fixed halo mass are brighter at higher redshifts, due to increased mass accretion rates (e.g., Mason et al. 2015).

In the right panel of Figure 2 we map our measured UV magnitudes to the inferred halo mass, $M_h$, using the UV luminosity–halo mass model of Mason et al. (2015) to compare with the redshift-independent $\Delta v_{\mathrm{Ly}\alpha}(M_h)$ model of Mason et al. (2018a). When transforming to halo mass we see a similar trend as with $M_{\mathrm{UV}}$, with less massive galaxies having lower Ly$\alpha$ offsets, and that our new results are consistent with the model, suggesting there is not strong evolution of $\Delta v_{\mathrm{Ly}\alpha}(M_h)$ with redshift. While our new data follow the model trend, our sample appears to have slightly higher velocity offsets than predicted at faint UV luminosities, which we discuss further in Section 5.1.

We see reasonable agreement with the models for galaxies between $6 < z < 7.5$ from the literature. Galaxies with $z > 7.5$ (Stark et al. 2017; Bunker et al. 2023; Tang et al. 2023) have much higher Ly$\alpha$ offsets than those expected for galaxies given their $M_{\mathrm{UV}}$ or halo mass, indicating a possible bias toward detecting high Ly$\alpha$ offsets during the epoch of reionization, which we will discuss in Section 5.2.

### 4.2. $\Delta v_{Ly\alpha}$ and Nebular Line Velocity Dispersion

The emerging Ly$\alpha$ line profile is expected to be shaped by both the column density ($N_{\mathrm{HI}}$) and the kinematics of the neutral gas close to line center (e.g., Verhamme et al. 2006, 2008; Hashimoto et al. 2015; Henry et al. 2015; Kakiichi & Gronke 2021; Hayes et al. 2023). As Ly$\alpha$ and H$\alpha$ likely are produced in the same H II regions, we therefore expect the Ly$\alpha$ profile may be linked to the kinematics of the gas as traced by H$\alpha$ (e.g., see Hayes et al. 2023, who find the correlation between optical line dispersion and $\Delta v_{\mathrm{Ly}\alpha}$ is primarily driven by the correlation of both of these properties with mass). We follow Erb et al. (2014) and examine the correlation between $\Delta v_{\mathrm{Ly}\alpha}$ and the velocity dispersion of the optical lines, as measured by rest-frame optical emission lines. We report the intrinsic dispersion, corrected for the instrumental broadening due to the resolution of our NIRSpec data, as $\sigma_{\mathrm{opt,intr}} = \sqrt{\sigma_{\mathrm{obs}}^2 - \sigma_{\mathrm{res}}^2}$, with $\sigma_{\mathrm{res}} \approx 47$ km s$^{-1}$.

In Figure 3 we plot $\Delta v_{\mathrm{Ly}\alpha}$ and $\sigma_{\mathrm{opt}}$ for our sample and compare with the Erb et al. (2014) sample. The majority of our sample has very a low intrinsic velocity dispersion, $\lesssim 50$ km s$^{-1}$, implying low dynamical masses. The literature sample shows a positive correlation between $\Delta v_{\mathrm{Ly}\alpha}$ and $\sigma_{\mathrm{opt}}$ and our new sample extends it to higher redshift, showing no strong evolution.

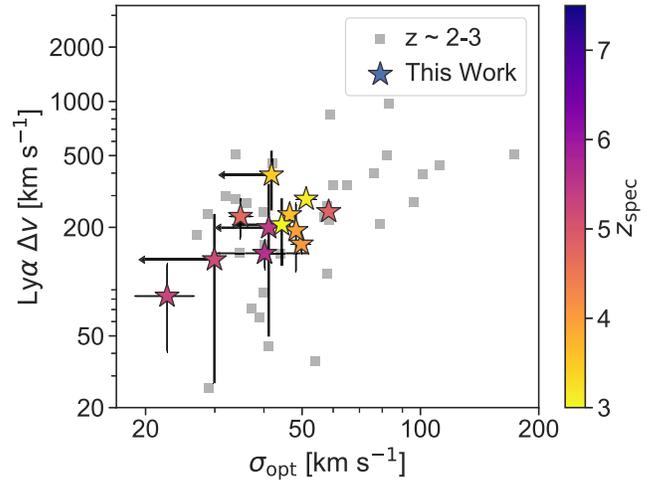

**Figure 3.** $\Delta v_{\mathrm{Ly}\alpha}$ vs. velocity dispersion of the optical lines. We present our data as stars color coded by redshift. In gray squares are $z \sim 2$–3 sources from Erb et al. (2014). We show upper limits for measurements below $\sigma_{\mathrm{res}} = 47$ km s$^{-1}$. Uncertainties are dominated by the MUSE Ly$\alpha$ centroid measurements. Both samples suggest a correlation between the optical line velocity dispersions and $\Delta v_{\mathrm{Ly}\alpha}$.

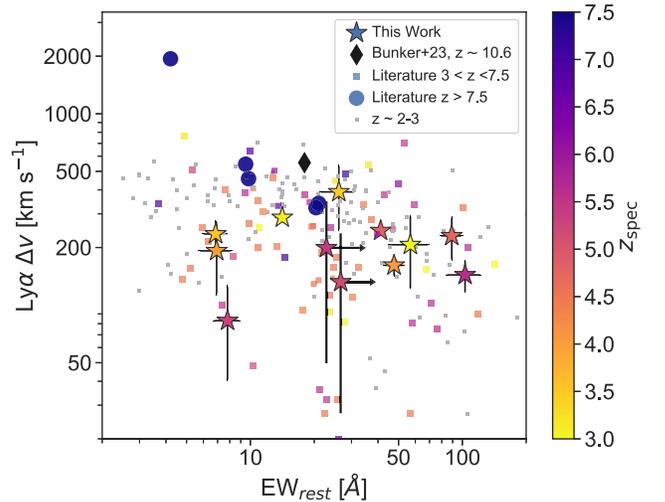

**Figure 4.** $\Delta v_{\mathrm{Ly}\alpha}$ against rest-frame EW of Ly$\alpha$. Our data are shown as stars color coded by redshift. We present values from the literature at $z \sim 2$–3 (Erb et al. 2014; Steidel et al. 2014; gray squares), at $3 < z < 7.5$ (Bradač et al. 2017; Stark et al. 2015; Willott et al. 2015; Inoue et al. 2016; Pentericci et al. 2016; Stark et al. 2017; Mainali et al. 2018; Cassata et al. 2020; Tang et al. 2021; Endsley et al. 2022; colored squares), $z > 7.5$ (Stark et al. 2017; Tang et al. 2023; blue circles), and GN-z11 (Bunker et al. 2023; black diamond). Both the literature data and our sample show low $\Delta v_{\mathrm{Ly}\alpha}$ at higher EW.

### 4.3. $\Delta v_{Ly\alpha}$ and EW

Previous works have found an anticorrelation between $\Delta v_{\mathrm{Ly}\alpha}$ and the Ly$\alpha$ EW (e.g., Erb et al. 2014; Tang et al. 2021). Both of these quantities are correlated with the neutral gas column density and scattering events (Verhamme et al. 2006, 2008; Yang et al. 2017a). Higher $N_{\mathrm{HI}}$, or a high covering fraction of neutral gas, makes it less likely for a Ly$\alpha$ photon to escape at its central wavelength, increasing $\Delta v_{\mathrm{Ly}\alpha}$; it can also imply higher dust content (Santini et al. 2014), and thus decreasing Ly$\alpha$ EW (Charlot & Fall 1993).

In Figure 4 we can see these effects for sources in the literature at $z \sim 2$–3 (Erb et al. 2014; Steidel et al. 2014); $3 < z < 7.5$ (Stark et al. 2015, 2017; Willott et al. 2015;

Inoue et al. 2016; Pentericci et al. 2016; Bradač et al. 2017; Mainali et al. 2018; Cassata et al. 2020; Tang et al. 2021; Endsley et al. 2022); and $z > 7.5$ (Stark et al. 2017; Tang et al. 2023), where galaxies with low Ly$\alpha$ EWs more commonly have high Ly$\alpha$ offsets. In general we find our sources lie in the same region as the literature sources. However, the bulk of galaxies in our sample have $\Delta v_{\mathrm{Ly}\alpha} < 300$ km s$^{-1}$, considerably lower than many sources in the literature. This is not surprising, since our sample only covers a small and less massive range of halo masses, where high $\Delta v_{\mathrm{Ly}\alpha}$ is less common. For Ly$\alpha$ EWs $< 20$ Å, all of our sample have Ly$\alpha$ velocity offsets below 300 km s$^{-1}$, lower than the high-redshift sources in the literature ($z > 7.5$).

One source, 70018, has both a low Ly$\alpha$ EW $= 8 \pm 2$ Å and low $\Delta v_{\mathrm{Ly}\alpha} = 88 \pm 42$, which is rare compared to the literature and physically unexpected, albeit the $\Delta v_{\mathrm{Ly}\alpha}$ error is high, and the source is only $\sim 2.5\sigma$ away from galaxies at similar EWs. We find that the galaxy is formed of two clumps in the UV and optical continuum, but only the main clump has H$\alpha$ and Ly$\alpha$ emissions, likely indicating a younger stellar population. The low Ly$\alpha$ EW and velocity offsets can be explained by dust absorption (Laursen et al. 2009). However, the galaxy does not appear to be particularly reddened ($\beta = -2.1$) and Roy et al. (2023) measure a low Balmer decrement H$\alpha$/H$_\beta = 2.89$. Nevertheless, Roy et al. (2023) measure a low Ly$\alpha$ escape fraction of 3% implying significant Ly$\alpha$ absorption in the ISM. The lack of emission lines in the second clump is not enough to result in a low EW due to it being fainter in the UV than the emitting clump. We find no evidence of a strong magnification gradient in the galaxy when using the lens models from Bergamini et al. (2023), which could cause differential magnification and strongly effect Ly$\alpha$ flux measurements.

## 5. Discussion

The results described in Section 4 provide a qualitative framework for understanding the transmission of Ly$\alpha$ through the ISM and through the reionizing IGM. We discuss the physical picture of Ly$\alpha$ scattering in the ISM emerging from our results in Section 5.1 and the implications of our results for Ly$\alpha$ visibility from UV-faint galaxies in the epoch of reionization in Section 5.2.

### 5.1. Physical Drivers of Ly$\alpha$ Velocity Offsets

The shift of Ly$\alpha$ photons to wavelengths redward and/or blueward of the central wavelength is due to resonant scattering with neutral hydrogen in the ISM and absorption by dust (e.g., Neufeld 1991; Verhamme et al. 2006). The amplitude of the velocity shift will be linked to the number of scattering events, which is set by the column density of neutral gas, $N_{\mathrm{HI}}$ (e.g., Neufeld 1990; Verhamme et al. 2006; Hashimoto et al. 2015; Henry et al. 2015; Kakiichi & Gronke 2021), its covering fraction (e.g., Shibuya et al. 2014; Jaskot et al. 2019), and the velocity of H I gas and any outflows (e.g., Pettini et al. 2001; Shapley et al. 2003).

In Figure 2 we demonstrate that data from a range of redshifts imply a correlation between $\Delta v_{\mathrm{Ly}\alpha}$ and UV luminosity, and therefore halo mass. This correlation is likely due to both to higher $N_{\mathrm{HI}}$ in more massive galaxies and increased velocity of H I gas in the ISM and higher outflow velocities (e.g., Heckman et al. 2015; Xu et al. 2022). We also find that several of our UV-faint galaxies show double-peaked Ly$\alpha$, implying a line of sight with lower $N_{\mathrm{HI}}$ and thus may be Lyman-continuum-leaker candidates (Roy et al. 2023).

While our results demonstrate a decrease in the Ly$\alpha$ velocity offset with decreasing UV luminosity our results appear to lie above the Mason et al. (2018a) model predictions, especially at the lowest $M_{\mathrm{UV}}$. We note that the model is semiempirical and was made using observations of $z \sim 2$ galaxies brighter than $M_{\mathrm{UV}} \sim -19$, thus the model lines fainter than this point in Figure 2 are an extrapolation both in redshift and UV luminosity (we also note the model predicts a large intrinsic scatter in $\Delta v_{\mathrm{Ly}\alpha}$—see Figure 2 in Mason et al. 2018a). Below we discuss a couple of physical factors which could cause our $z \sim 3$–6 sample to have higher velocity offsets than the model predictions.

It is possible that our $z \sim 3$–5.6 sample is biased toward higher $\Delta v_{\mathrm{Ly}\alpha}$, compared to the $z \sim 2$ samples the model is based on, due to the increasing optical depth in the IGM (Gunn & Peterson 1965) with increasing redshift (e.g., Faucher-Giguère et al. 2008; Becker et al. 2015). In particular, the infall of dense gas around dark matter halos may lower the transmission of Ly$\alpha$ emitted redward of systemic, as red photons are blueshifted into the Ly$\alpha$ resonant frequency in the frame of the infalling gas (e.g., Santos 2004; Dijkstra et al. 2007; Laursen et al. 2011). We test the expected impact of this scenario by using Ly$\alpha$ transmission curves extracted from the IllustrisTNG-100 simulations (Nelson et al. 2019) of Byrohl et al. (2019). We use 5000 randomly selected mass halos and lines of sight. At $z \sim 5$ we find Ly$\alpha$ transmission rises above 30% at a median $\Delta v_{\mathrm{Ly}\alpha}$ of 90 km s$^{-1}$ (with a standard deviation of 50 km s$^{-1}$), making detections of Ly$\alpha$ with velocity offsets lower than this rare. By contrast, at $z \sim 2$, where the model of Mason et al. (2018a) was calibrated, the median $\Delta v_{\mathrm{Ly}\alpha}$ where $>30\%$ Ly$\alpha$ is transmitted decreases to 60 km s$^{-1}$, making it easier to detect low $\Delta v_{\mathrm{Ly}\alpha}$. The impact of resonant absorption in the IGM could be tested by seeing how both the velocity offset and spectral shape of Ly$\alpha$ evolves with redshift, as resonant absorption due to infalling will cause a sharp cutoff on the "blue side" of the Ly$\alpha$ line redward of systemic (e.g., Park et al. 2021).

Observing higher $\Delta v_{\mathrm{Ly}\alpha}$ than the models could also be achieved with an increased presence of outflows with increasing redshift; for example, due to increased star formation rates at fixed UV luminosity with increasing redshift. Ly$\alpha$ photons scatter in outflowing gas, making it preferentially easier to detect photons which are "backscattered" on the far side of the source to the observer. These photons are Doppler shifted to a velocity offset $\sim 2\times$ the outflow velocity and thus the photons fall out of resonance and transmit through the gas more easily, making redshifted Ly$\alpha$ line profiles ubiquitous in galaxies with strong outflows (e.g., Shapley et al. 2003; Verhamme et al. 2006; Erb et al. 2014). The incidence and velocity of outflows at $z > 2$ can now be measured more easily with JWST (see, e.g., Carniani et al. 2023).

In Figure 3, we see a correlation between $\Delta v_{\mathrm{Ly}\alpha}$ and the velocity dispersion of optical lines, finding that our sample of galaxies with low $\Delta v_{\mathrm{Ly}\alpha}$ also have low optical line velocity dispersions. This suggests that Ly$\alpha$ scattering is correlated to the velocity dispersion of gas close to the line center, as suggested by Erb et al. (2014). Since the nebular emission lines we observed, H$\alpha$ and [O III], are produced in the same H II regions around massive stars as Ly$\alpha$, and are not affected by resonant scattering, but broadened by the velocity distribution of the ionized gas, their line profiles should contain information about the velocity dispersion of the H II regions. At the same time, $\Delta v_{Ly\alpha}$ and Ly$\alpha$ broadening is produced by a combination of the column density and kinematics of H I (resonant scattering). All of the aforementioned processes are correlated with galaxy mass, as shown in the analysis of Hayes et al. (2023) for $z \sim 0$–0.4 LAEs, who find that the correlation of optical line dispersion with $\Delta v_{Ly\alpha}$ is mostly driven by the correlation of both quantities with mass. This is consistent with the picture proposed by Henry et al. (2015) that H I column density dominates Ly$\alpha$ escape in low-mass galaxies.

### 5.2. Implications for Ly$\alpha$ Observations in the Epoch of Reionization

One of the main results we can conclude from Figure 2, is that UV-faint galaxies ($M_{UV} < -20$) at $z \sim 3$–6 tend to have low $\Delta v_{Ly\alpha}$ ($\lesssim 300$ km s$^{-1}$). As the galaxy population shifts to faint luminosities at high redshift (e.g., Bouwens et al. 2021) we expect that a large fraction of galaxies at $z \gtrsim 6$ may be UV-faint galaxies emitting Ly$\alpha$ with low velocity offsets and intrinsically high Ly$\alpha$ EWs ($\gtrsim 25$ Å; e.g., Stark et al. 2011; Cassata et al. 2015; De Barros et al. 2017; Fuller et al. 2020). When the IGM was still significantly neutral, as during and before the epoch of reionization ($z \sim 6$–10; e.g., Mason et al. 2019), we expect the Ly$\alpha$ damping wing to attenuate Ly$\alpha$ photons that escape the galaxies strongly or completely near their central wavelength (e.g., Dijkstra et al. 2011; Mason & Gronke 2020).

In this case it would be possible for Ly$\alpha$ to propagate through the IGM to the observer only if it has a high enough $\Delta v_{Ly\alpha}$ such that the photons do not experience the strong damping wing. Assuming that the line is not broadened and weakened out of the detection limit of the observations, this should then produce a bias, where we will be more likely to see galaxies with very high $\Delta v_{Ly\alpha}$ ($\gtrsim 500$ km s$^{-1}$) during the epoch of reionization, even if such sources are less common in the population. We can see hints of this in the results of Stark et al. (2017) and Tang et al. (2023): they report Ly$\alpha$ emission from UV-bright ($M_{UV} \lesssim -21$) galaxies at $z > 7.5$, well within the epoch of reionization. All of these galaxies have $\Delta v_{Ly\alpha} > 300$ km s$^{-1}$, higher than most of our sample, and higher than the predictions of Mason et al. (2018a). The results from Bunker et al. (2023) show that this trend continues at higher redshift in a detected Ly$\alpha$ galaxy at $z \sim 10.6$, with $\Delta v_{Ly\alpha} = 555$ km s$^{-1}$.

This is in contrast to the low detection rates of Ly$\alpha$ in UV-faint galaxies at $z \gtrsim 7$ (Hoag et al. 2019a; Mason et al. 2019), including the case of A2744-z7p9OD, a $z = 7.88$ spectroscopically confirmed overdensity where none of the seven UV-faint ($M_{UV} > -20$) galaxies in a 60 kpc proper radius have observed Ly$\alpha$ emission > 16–28 Å (Morishita et al. 2023). Our results imply that the low Ly$\alpha$ velocity offsets in UV-faint galaxies likely contribute to the preferential attenuation of Ly$\alpha$ from these galaxies during reionization, in addition to these galaxies likely tracing less overdense and late reionized regions of the IGM compared to UV-bright galaxies (Mason et al. 2018b; Lu et al. 2023).

The detection of Ly$\alpha$ with low $\Delta v_{Ly\alpha}$ in reionization epoch galaxies can be an important indicator of the presence and size of large ionized bubbles; Ly$\alpha$ emitted from galaxies inside large ionized bubbles does not need to have have been scattered far from its central wavelength in order to transmit through the IGM. In Figure 5 we illustrate this by showing the fraction of Ly$\alpha$ flux transmitted through the IGM, assuming a single ionized bubble in a neutral IGM, following Mason & Gronke (2020), as a function of ionized bubble size and emitted $\Delta v_{Ly\alpha}$. Here we make mock Ly$\alpha$ emission lines with FWHM = $\Delta v_{Ly\alpha}$ (Verhamme et al. 2015; Mason et al. 2018a; fixing FWHM = 50 km s$^{-1}$ for lines with $\Delta v_{Ly\alpha} < 50$ km s$^{-1}$) and calculate the transmission fraction as the ratio of total observed flux to total emitted flux. This shows that the detection of Ly$\alpha$ at $\lesssim 200$ km s$^{-1}$ with a high Ly$\alpha$ escape fraction, implying significant transmitted flux close to line center, provides strong evidence of $\gtrsim 1$ proper Mpc (pMpc) ionized bubbles (see Saxena et al. 2023 for a recent candidate), but that Ly$\alpha$ with $\Delta v_{Ly\alpha} \gtrsim 400$ km s$^{-1}$ can still transmit $\gtrsim 10\%$ flux through the IGM even if the source is inside a very small ionized bubble (or even in a fully neutral IGM; see also Dijkstra et al. 2011; Mason & Gronke 2020; Endsley et al. 2022). UV-faint galaxies around UV-bright Ly$\alpha$-emitting galaxies are thus ideal targets for deep spectroscopic follow up to confirm and estimate the sizes of ionized bubbles, as they have high number densities and are likely to emit their Ly$\alpha$ at low velocity offsets from systemic. In the case where we detect no Ly$\alpha$ from the surrounding faint galaxies then the reionization bubble is likely not sufficiently large to allow Ly$\alpha$ transmission close to the line center.

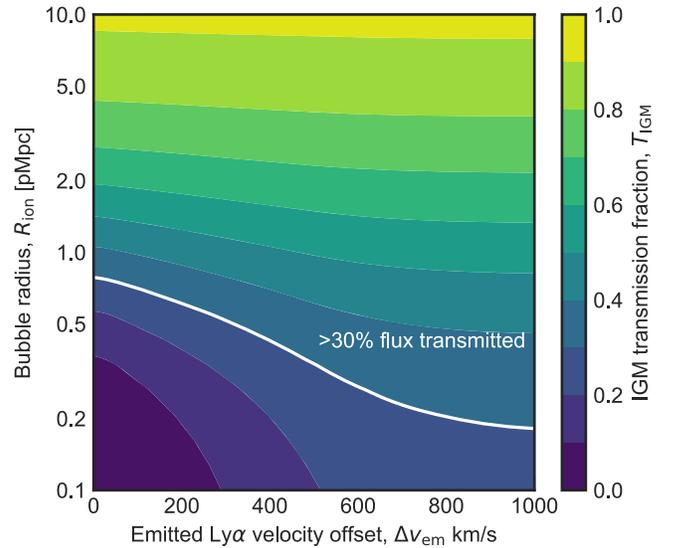

**Figure 5.** Fraction of Ly$\alpha$ flux transmitted through the IGM as a function of the radius of the ionized bubble and emitted $\Delta v_{Ly\alpha}$, color coded by transmission through the IGM. We show the contour of 30% transmission as a white curve. The IGM transmission values are calculated assuming a Ly$\alpha$-emitting galaxy is in the center of a single ionized bubble in a neutral IGM, as described in Section 5.2. This figure shows that Ly$\alpha$ emission lines with $\Delta v_{Ly\alpha} \lesssim 200$ km s$^{-1}$ must reside at least $\gtrsim 1$ pMpc from the neutral IGM in order to transmit a significant fraction of their flux through the IGM.

Furthermore, it may prove useful to use the properties of optical emission lines, such as the correlation between $\Delta v_{Ly\alpha}$ and $\sigma_{opt}$ to estimate the *emitted* Ly$\alpha$ line profile (before attenuation in the IGM) and thus infer the attenuation of Ly$\alpha$ during the epoch of reionization. While we observe a large scatter in the correlation between $\Delta v_{Ly\alpha}$ and $\sigma_{opt}$, JWST enables optical line spectroscopy for large samples of $z \sim 3$–6 LAEs and thus opens the door to finding good estimators of emitted Ly$\alpha$ line profiles (e.g., Hayes et al. 2023).

## 6. Conclusions

We have measured the velocity offset of Lyα emission from systemic in 12 galaxies at $z \sim 3-6$, as well as the velocity dispersion of their optical emission lines, Hα and [O III] λ5007. Thanks to JWST and the lensed cluster A2744, we were able to measure $\Delta v_{Ly\alpha}$ at UV magnitudes not previously possible ($-19 \lesssim M_{UV} \lesssim -16$). Our conclusions are as follows:

1. Our sample has a median $M_{UV}$ of $-17.8 \pm 1.4$ and we find a median and $1\sigma$ dispersion $\Delta v_{Ly\alpha} = 205 \pm 75$ km s$^{-1}$. Combined with observations from previous work targeting more luminous galaxies we find a positive correlation between UV luminosity and $\Delta v_{Ly\alpha}$, consistent with a theoretical model where $\Delta v_{Ly\alpha}$ depends on halo mass.

2. We observe that our sample follows the previously reported correlation between $\Delta v_{Ly\alpha}$ and the velocity dispersion of optical lines, though with a large scatter. With upcoming large samples made with JWST, properties of optical emission lines could prove to be a useful predictor of Lyα line properties during the epoch of reionization.

3. We find agreement with the literature when comparing $\Delta v_{Ly\alpha}$ and Lyα EW, with the majority of sources falling on a trend of increasing EW with decreasing $\Delta v_{Ly\alpha}$.

4. Our results are consistent with, and extend toward fainter magnitudes, the framework explored by previous studies that Lyα velocity offsets are driven by the abundance and velocity of neutral hydrogen close to the H II regions where Lyα is produced. Our results are thus consistent with the assumption that Lyα emission from UV-faint galaxies during the epoch of reionization may be preferentially attenuated compared to that from UV-bright galaxies because of the strong damping wing optical depth close to line center produced by intergalactic neutral hydrogen.

We have demonstrated JWST/NIRSpec is a powerful tool to understand the properties of high-redshift galaxies and to help in our understanding of the reionization era. An exciting prospect for the JWST era is the robust confirmation of ionized bubbles at $z \gtrsim 6$. Observations of UV-faint galaxies with low Lyα velocity offsets should indicate the presence of large ionized bubbles, as to observe Lyα close to systemic velocity during reionization implies a low optical depth at line center. The demonstrated low Lyα velocity offsets in our sample motivate deep, high-resolution spectroscopy of reionization-era galaxies with JWST/NIRSpec, to obtain Lyα and optical line measurements to enable such studies.

### Acknowledgments

This work is based on observations on observations made with the NASA/ESA/CSA James Webb Space Telescope. The JWST data were obtained from the Mikulski Archive for Space Telescopes at the Space Telescope Science Institute, which is operated by the Association of Universities for Research in Astronomy, Inc., under NASA contract NAS 5-03127 for JWST. These observations are associated with program JWST-ERS-1324. The specific observations analyzed can be accessed via doi:10.17909/9a2g-sj78. G.P.L. and C.M. acknowledge support from the VILLUM FONDEN under grant 37459. The Cosmic Dawn Center (DAWN) is funded by the Danish National Research Foundation under grant DNRF140. We acknowledge financial support from NASA through grant JWST-ERS-1324. M.B. acknowledges support from the Slovenian national research agency ARRS through grant N1-0238. We thank Dawn Erb and Dan Stark for sharing velocity offset measurements.

*Facilities*: VLT (MUSE), JWST (NIRSpec), and HST (ACS).

### ORCID iDs

Gonzalo Prieto-Lyon https://orcid.org/0000-0003-3518-0374
Charlotte Mason https://orcid.org/0000-0002-3407-1785
Sara Mascia https://orcid.org/0000-0002-9572-7813
Emiliano Merlin https://orcid.org/0000-0001-6870-8900
Namrata Roy https://orcid.org/0000-0002-4430-8846
Alaina Henry https://orcid.org/0000-0002-6586-4446
Guido Roberts-Borsani https://orcid.org/0000-0002-4140-1367
Takahiro Morishita https://orcid.org/0000-0002-8512-1404
Xin Wang https://orcid.org/0000-0002-9373-3865
Kit Boyett https://orcid.org/0000-0003-4109-304X
Patricia Bolan https://orcid.org/0000-0002-7365-4131
Marusa Bradač https://orcid.org/0000-0001-5984-0395
Marco Castellano https://orcid.org/0000-0001-9875-8263
Amata Mercurio https://orcid.org/0000-0001-9261-7849
Themiya Nanayakkara https://orcid.org/0000-0003-2804-0648
Diego Paris https://orcid.org/0000-0002-7409-8114
Laura Pentericci https://orcid.org/0000-0001-8940-6768
Claudia Scarlata https://orcid.org/0000-0002-9136-8876
Michele Trenti https://orcid.org/0000-0001-9391-305X
Tommaso Treu https://orcid.org/0000-0002-8460-0390
Eros Vanzella https://orcid.org/0000-0002-5057-135X